\documentclass[twocolumn,aps,pra,showpacs,floatfix,groupedaddress]{revtex4}
\usepackage{times}
\usepackage{graphicx}
\usepackage{amsmath}
\usepackage{amssymb}
\usepackage{hyperref}
\usepackage{latexsym}
\usepackage{epsfig}


\begin{document}

\title{Heteronuclear molecules in an optical lattice: Theory and experiment}

\author{F. Deuretzbacher, K. Plassmeier, D. Pfannkuche}
\affiliation{I. Institut f\"ur Theoretische Physik, Universit\"at Hamburg, Jungiusstra{\ss}e 9, 20355 Hamburg, Germany}
\author{F. Werner}
\affiliation{Laboratoire Kastler Brossel, ENS, UPMC, CNRS, 24 rue Lhomond, 75231 Paris Cedex 05, France}
\author{C. Ospelkaus$^1$, S. Ospelkaus$^1$, K. Sengstock$^1$, K. Bongs$^{1,2}$}
\affiliation{$^1$Institut f\"ur Laserphysik, Universit\"at Hamburg, Luruper Chaussee 149, 22761 Hamburg, Germany \\ $^2$Midlands Centre for Ultracold Atoms, School of Physics and Astronomy, University of Birmingham, Edgbaston, Birmingham B15 2TT, United Kingdom}

\begin{abstract}
We study properties of two different atoms at a single optical lattice site at a heteronuclear atomic Feshbach resonance. We calculate the energy spectrum, the efficiency of rf association and the lifetime as a function of magnetic field and compare the results with the experimental data obtained for $^{40}$K and $^{87}$Rb [C. Ospelkaus {\it et al.}, Phys. Rev. Lett. {\bf 97}, 120402 (2006)]. We treat the interaction in terms of a regularized $\delta$ function pseudopotential and consider the general case of particles with different trap frequencies, where the usual approach of separating center-of-mass and relative motion fails. We develop an exact diagonalization approach to the coupling between center-of-mass and relative motion and numerically determine the spectrum of the system. At the same time, our approach allows us to treat the anharmonicity of the lattice potential exactly. Within the pseudopotential model, the center of the Feshbach resonance can be precisely determined from the experimental data.
\end{abstract}


\maketitle

\section*{\uppercase{Introduction}}

\vspace{-2ex}

Motivated by the intriguing perspectives of heteronuclear molecule formation, observation of charge-density waves~\cite{CDW}, boson-induced fermionic superfluidity~\cite{BIndCoop1,BIndCoop2,BIndCoop3} in optical lattices~\cite{BIndCoop4}, and supersolids~\cite{supersolid}, Fermi-Bose mixtures have recently attracted lots of attention. An important step in this direction was the simultaneous trapping of bosons and fermions in a three-dimensional (3D) optical lattice~\cite{ImpIndLoc,FB3DLat}. Recently, even heteronuclear molecules~\cite{HeteronuclearMolecules,Papp06} were created by means of a magnetic field Feshbach resonance in combination with rf association~\cite{HeteronuclearMolecules}.

In interpreting the experimental results and for future extensions, it is essential to develop a detailed understanding of the interaction of two particles across the Feshbach resonance, taking into account the external confinement of the optical lattice in a consistent manner. In a seminal paper, Th.~Busch {\it et al.}~\cite{Busch98} have analytically solved the problem of two identical atoms in a harmonic trap. This model has been compared to two-component Fermi gases in an optical lattice at a Feshbach resonance~\cite{ETHLatticeMol,ETH_langes_Paper}.

In this paper, we study the generalized case of two different atoms at an optical lattice site accounting for the anharmonic part of the potential. Both the fact that the two atoms feel different trap frequencies and the anharmonicity lead to a coupling of center-of-mass and relative motion of the two atoms resulting in deviations from the model in Ref.~\cite{Busch98}. We model interactions between two cold atoms by a regularized $\delta$ function type interatomic pseudopotential. We discuss the solutions of the uncoupled problem and develop an exact diagonalization approach to the coupling term. In this very general approach, we find significant deviations from the identical particle scenario, the strongest effect being observed for repulsively interacting atoms with large mass ratios. We further discuss rf association as a method of determining the energy spectrum at a heteronuclear Feshbach resonance between $^{87}$Rb and $^{40}$K in a 3D optical lattice. We compare the theoretical energy spectrum to the experimental results~\cite{HeteronuclearMolecules} and discuss methods of precisely determining the Feshbach resonance center position based on this comparison. Finally we calculate the efficiency of rf association and the lifetime of heteronuclear $^{40}$K-$^{87}$Rb molecules and find qualitative agreement with experimental results.

\vspace{-3ex}

\section{Theoretical model}

\vspace{-2ex}

In order to model interactions within an atom pair, we consider an interatomic potential given by a regularized $\delta$ potential~\cite{validity,Castin,C6,SimoniPrivComm}. For two atoms of the same kind in an isotropic harmonic trap an analytic solution exists~\cite{Busch98}. Here we consider two different atomic species which are confined at a single site of a 3D optical lattice. In this case the atoms experience different trapping frequencies and the confining potential has significant anharmonic features. We use the following Hamiltonian as a starting point
\begin{eqnarray} \label{definition_H}
  H & = & \sum_{i=1,2} \left[ -\frac{\hbar^2}{2m_i} \Delta_i + \frac{1}{2} m_i \omega_i^2 r_i^2 \right] + \frac{2 \pi \hbar^2 a_s}{\mu} \delta(\vec r) \frac{\partial}{\partial r} r \nonumber \\
  & & + V_\mathrm{corr}(\vec r_1, \vec r_2) .
\end{eqnarray}
Here $m_1$ and $m_2$ are the masses of the two atoms, ${\omega_i = \sqrt{2V_i k^2/m_i}}$ are the trapping frequencies obtained using the harmonic approximation to the trapping potential, $k$ is the wave number and $V_i$ is the depth of the lattice felt by atom $i$, $a_s$ is the scattering length, ${\mu = m_1 \cdot m_2 / M}$ the reduced mass, ${M=m_1+m_2}$ the total mass, ${\vec{r}=\vec{r}_1-\vec{r}_2}$ the relative position, and ${r = |\vec r_1 - \vec r_2|}$ is the distance between the atoms. $V_\mathrm{corr}$ contains the anharmonic corrections which are necessary to accurately approximate the potential of one lattice site given by
\begin{equation}
  V_\mathrm{lattice} = V_\mathrm{lattice}^{(x)} + V_\mathrm{lattice}^{(y)} + V_\mathrm{lattice}^{(z)}
\end{equation}
with
\begin{eqnarray} \label{Lattice}
  V_\mathrm{lattice}^{(x)} & = & \sum_{i=1,2} V_i \sin^2(k x_i) \nonumber \\
  & \approx & \sum_{i=1,2} \left[ V_i k^2 x_i^2 -\frac{V_i k^4}{3} x_i^4 + \ldots \right]
\end{eqnarray}
and similar expressions for $V_\mathrm{lattice}^{(y)}$ and $V_\mathrm{lattice}^{(z)}$. The first term of Eq.~(\ref{Lattice}) gives rise to the harmonic approximation through $\omega_i=\sqrt{2V_ik^2/m_i}$, and the remainder gives rise to $V_\mathrm{corr}$.

We introduce relative and center-of-mass coordinates ${\vec r = \vec r_1 - \vec r_2}$ and $\vec R = (m_1 \vec r_1 + m_2 \vec r_2)/M$ and define the corresponding frequencies:
\begin{eqnarray}
  \omega_\mathrm{c.m.} & := & \sqrt{\frac{m_1 \omega_1^2 + m_2 \omega_2^2}{m_1 + m_2}},
  \\
  \omega_\mathrm{rel} & := & \sqrt{\frac{m_2 \omega_1^2 + m_1 \omega_2^2}{m_1 + m_2}},
  \\
  \Delta \omega & := & \sqrt{\omega_1^2 - \omega_2^2} .
\end{eqnarray}
The transformed Hamiltonian consists of three contributions: one center-of-mass harmonic oscillator Hamiltonian $H_\mathrm{c.m.}$, one term for the relative motion $H_\mathrm{rel}$ containing a harmonic oscillator term and the contact interaction, and one last term $H_\mathrm{couple}$ grouping together all terms which couple relative and center-of-mass motion and which arise from the different trap frequencies and the anharmonic corrections:
\begin{eqnarray}
\label{H}
  H & = & -\frac{\hbar^2}{2M} \Delta_\mathrm{c.m.} + \frac{1}{2} M \omega_\mathrm{c.m.}^2 R^2 - \frac{\hbar^2}{2\mu} \Delta_\mathrm{rel} + \frac{1}{2}\mu\omega_\mathrm{rel}^2 r^2 \nonumber \\
    & & +\frac{2 \pi \hbar^2 a_s}{\mu} \delta(\vec r)\frac{\partial}{\partial r}r + \mu \Delta\omega^2 \vec r \cdot \vec R + V_\mathrm{corr}(\vec R, \vec r) \nonumber \\
    & =: & H_\mathrm{c.m.} + H_\mathrm{rel} + H_\mathrm{couple} .
\end{eqnarray}
Let us first neglect the coupling terms $H_\mathrm{couple}$. In this case, the problem separates into relative and center-of-mass motion, with the center-of-mass motion given by harmonic oscillator wave functions. The Hamiltonian of the relative motion $H_\mathrm{rel}$ is solved analytically in Ref.~\cite{Busch98} and leads to the energy structure
\begin{equation}
  2 \frac{\Gamma \left[ -E_\mathrm{rel}/(2 \hbar \omega_\mathrm{rel}) + 3/4 \right]}{\Gamma \left[ -E_\mathrm{rel}/(2 \hbar \omega_\mathrm{rel}) + 1/4 \right]} = \frac{1}{(a_s / a_\mathrm{rel})} \quad (l = 0) ,
\end{equation}
with $a_\mathrm{rel} = \sqrt{\hbar/(\mu \omega_\mathrm{rel})}$. Only $l=0$ states are considered here, since these are the only ones affected by the regularized $\delta$ potential. The rest of the spectrum consists of $l\neq 0$ harmonic oscillator states with an energy independent of $a_s$. The corresponding eigenfunctions are given by
\begin{equation} \label{phi_busch}
  \phi(r;\nu) = \frac{A}{a_\mathrm{rel}^{3/2}} \Gamma(-\nu) U\left( -\nu, \frac{3}{2}; (r/a_\mathrm{rel})^2 \right) e^{-(r/a_\mathrm{rel})^2/2} .
\end{equation}
$A$ is a normalization constant which we determine numerically, ${U(a,b;z)}$ are the confluent hypergeometric functions and the non-integer indices $\nu$ are related to the energy by ${E_\mathrm{rel} = \hbar \omega_\mathrm{rel} \left( 2 \nu + \frac{3}{2} \right)}$.

\begin{figure}[t]
  \centering
  \includegraphics[width=.9\columnwidth]{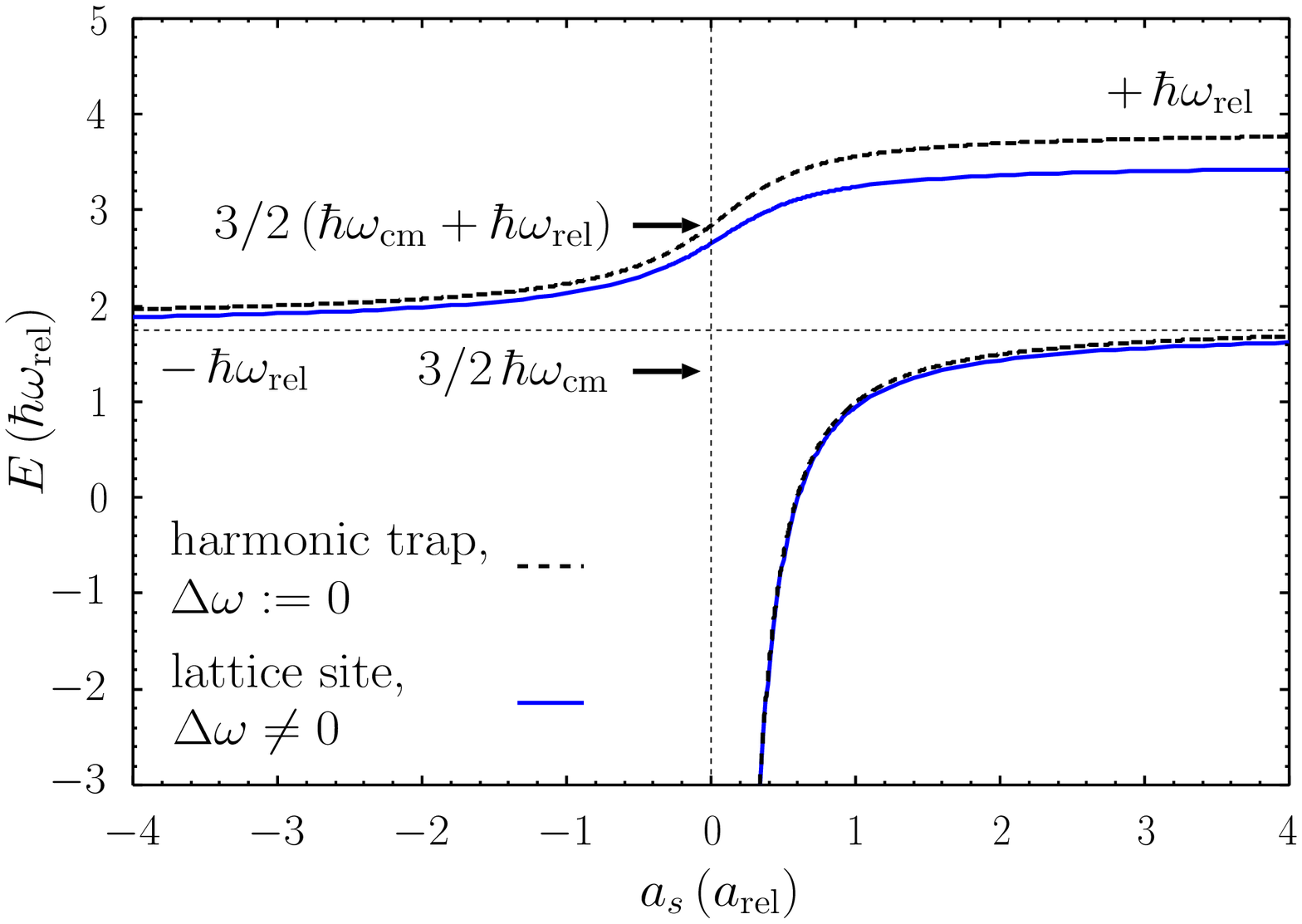}
  \caption{(Color online) Energy eigenvalues of $^{40}$K and $^{87}$Rb as a function of scattering length without (black dashed line) and with coupling terms (blue solid line) due to anharmonicity and unequal trap frequencies in the lattice for parameters: ${V_\mathrm{Rb} = 40.5 E_{r,\mathrm{Rb}}}$, ${V_\mathrm{K} = 0.86 V_\mathrm{Rb}}$, and {$\lambda = 1030$\,nm}. The deviation between the idealized model and the full solution is substantial in particular for the upper branch.}
\label{fig:eofa1}
\end{figure}

The resulting energy spectrum is shown in Fig.~\ref{fig:eofa1} (black dashed line) for a center-of-mass energy of $3/2 \, \hbar \omega_\mathrm{c.m.}$. For vanishing interaction, the lowest harmonic oscillator state has an energy of ${3/2(\hbar \omega_\mathrm{c.m.}+\hbar \omega_\mathrm{rel})}$. For large positive values of $a_s$, it transforms into repulsively interacting atom pairs with a unitary positive ``binding energy'' of $+\hbar\omega_\mathrm{rel}$. In a recent experiment with bosonic atoms in an optical lattice, such repulsively interacting atom pairs served as a starting point to create stable repulsively bound pairs~\cite{Innsbruck_pairs}. For negative $a_s$, the aforementioned state transforms into attractively interacting atoms. In the unitary limit ($a_s \rightarrow -\infty$), these atoms acquire a binding energy of $-\hbar\omega_\mathrm{rel}$. When the scattering length changes from large and negative to large and positive (as observed, e.~g., at atomic Feshbach resonances), we enter the molecule part of the spectrum. In that part of the spectrum, the resulting two-body bound state is stable even in the absence of the external potential. As $a_s$ becomes smaller and smaller again from above ($a_s\rightarrow +0$), the size of the molecule decreases proportionally to $a_s$, and
the binding energy tends to $-\infty$.

As soon as we add the coupling term, $H_\mathrm{couple}$, this treatment is no longer valid as center-of-mass and relative motion are no longer decoupled. In order to describe this problem in a consistent fashion, we have calculated the matrix of the complete Hamiltonian (\ref{H}) using the eigenfunctions of ${H_\mathrm{c.m.} + H_\mathrm{rel}}$ and numerically obtained energy eigenvalues and eigenfunctions for the coupled problem by diagonalizing $H$.

The anharmonic corrections are treated as follows. Since ${x_1 = X + a x}$ and ${x_2 = X - b x}$ with $a := m_2/M$ and $b := m_1/M$, the $x$-dependent part of the anharmonic corrections $V_\mathrm{corr}^{(x)}$ transforms to
\begin{eqnarray}
  V_\mathrm{corr}^{(x)} & \mspace{-9mu} = \mspace{-9mu} & -\frac{V_1 + V_2}{3} k^4 X^4 -\frac{4 (V_1 a - V_2 b)}{3} k^4 x X^3 \nonumber \\
  & & -2 (V_1 a^2 + V_2 b^2) k^4 x^2 X^2 - \frac{4 (V_1 a^3 - V_2 b^3)}{3} k^4 x^3 X \nonumber \\
  & & -\frac{V_1 a^4 + V_2 b^4}{3} k^4 x^4 + \ldots .
\end{eqnarray}
Corresponding expressions are obtained for the $y$- and $z$-direction, $V_\mathrm{corr}^{(y)}$ and $V_\mathrm{corr}^{(z)}$. In the numerical implementation, we have tested for convergence with terms up to eighth order. We found that including eighth-order corrections improve the accuracy of the calculation by only ${\approx 3 \times 10^{-3} \, \hbar \omega_\mathrm{rel}}$.

Our approach leads to the diagonalization of rather small Hamiltonian matrices as our main interest is the modification of the ground state and the repulsively interacting pair branch. The whole calculation has been done with {\footnotesize MATHEMATICA}. As basis we have chosen the states ${|N,L,M,n,l,m\rangle}$ with lowest principal quantum numbers ${\Pi := 2N+L+2n+l = 0, 1, ..., \Pi_\mathrm{max}}$, where $N$, $L$, $M$ and $n$, $l$, $m$ are the quantum numbers of the eigenfunctions of the rotationally symmetric harmonic oscillator of center-of-mass and relative motion, respectively. We typically used ${\Pi_\mathrm{max} = 7}$ leading to a total number of 258 basis states. We have found that adding another level of the uncoupled problem to the basis set leads to additional changes in the energy smaller than ${\approx 10^{-3} \, \hbar \omega_\mathrm{rel}}$. Furthermore, we exploited the fact that the total angular momentum ${L_z = \hbar (M + m)}$ of the low-energy eigenfunctions is approximately conserved despite the cubic symmetry of the optical lattice. Again, we found that including ${L_z \neq 0}$ basis states lowers the energy by only ${\approx 3 \times 10^{-3} \, \hbar \omega_\mathrm{rel}}$~\cite{Basis}.

Fig.~\ref{fig:eofa1} shows the resulting energy spectrum (blue solid line) compared to the uncoupled solution (black dashed line), calculated for $^{40}$K and $^{87}$Rb with the experimental parameters of Ref.~\cite{HeteronuclearMolecules}: ${V_\mathrm{Rb} = 40.5 \, E_{r,\mathrm{Rb}}}$, ${V_\mathrm{K} = 0.86 \, V_\mathrm{Rb}}$, and ${\lambda = 1030 \, \mathrm{nm}}$. ${E_{r,\mathrm{Rb}} = \hbar^2 k^2 / 2m_\mathrm{Rb}}$ is the $^{87}$Rb recoil energy. In the case of heteronuclear atom pairs it is useful to express the lattice depth in units of ${E_{r,\mathrm{rel}} = \hbar^2 k^2 / 2 \mu}$, which is the kinetic energy given to a particle with reduced mass $\mu$ by a photon of momentum ${\hbar k}$. Then, ${V_\mathrm{Rb} = 40.5 \, E_{r,\mathrm{Rb}} = 12.6 \, E_{r,\mathrm{rel}}}$. As can be seen from the figure, the deviation between the idealized model, where the coupling term has been neglected, and the full solution is substantial. The difference is most pronounced in the repulsively interacting pair branch (${0.34 \, \hbar \omega_\mathrm{rel} \approx 20}$\% of the level spacing) and becomes smaller as we enter the attractively interacting atom part of the spectrum. The molecular branch is relatively unaffected by the coupling term $H_\mathrm{couple}$. This is natural because as we approach $a\rightarrow +0$, the role of the external confinement decreases since the molecule becomes smaller.

\begin{figure}[t]
  \centering
  \includegraphics[width=.9\columnwidth]{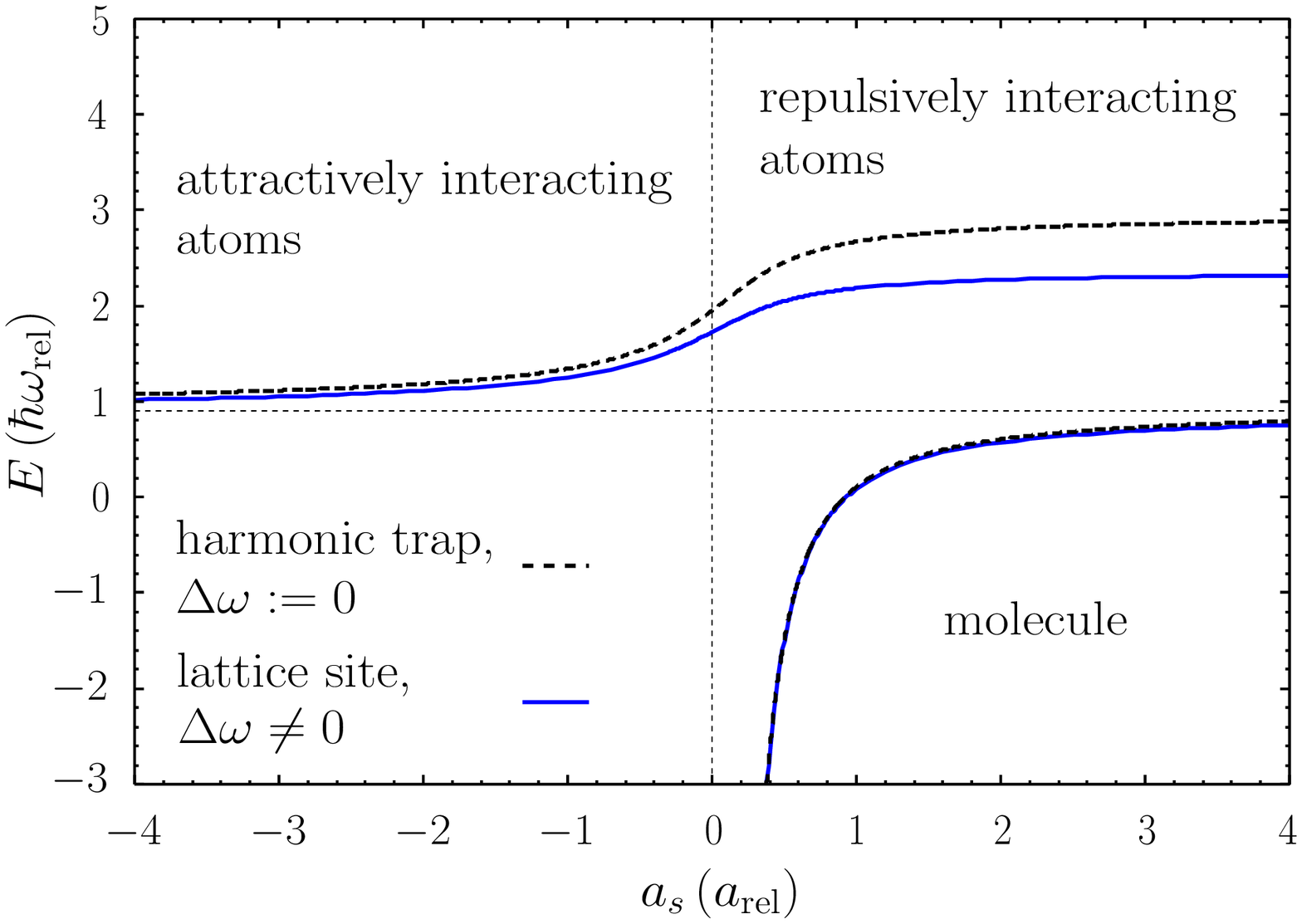}
  \caption{(Color online) Low-energy spectrum of states with center-of-mass energy $3/2 \, \hbar \omega_\mathrm{c.m.}$ for $^6$Li and $^{133}$Cs and lattice parameters ${V_\mathrm{Li} = V_\mathrm{Cs} = 10 \, \hbar^2 k^2 / 2 \mu}$ and {$\lambda = 1 \mu$m}. The energy is much more lowered compared to the case of $^{40}$K and $^{87}$Rb. This is due to the large mass ratio of the $^6$Li and $^{133}$Cs atoms.}
\label{fig:eofa2}
\end{figure}

The influence of the coupling term $H_\mathrm{couple}$ is even stronger if we consider molecules with large mass ratios as in the case of $^6$Li and $^{133}$Cs, see Fig.~\ref{fig:eofa2}. We have chosen the lattice parameters ${V_\mathrm{Li} = V_\mathrm{Cs} = 10 \, E_{r,\mathrm{rel}}}$ and ${\lambda = 1 \, \mu\mathrm{m}}$. Here the energy of the repulsively interacting atoms is lowered by up to ${\approx 0.6 \, \hbar \omega_\mathrm{rel}}$.

\begin{table}[b]
  \caption{\label{tab:table1} Influence of the individual coupling terms $H_{\Delta \omega}$ and $V_\mathrm{corr}$, onto the total energy of several atom pairs. The energies are given in units of ${\hbar \omega_\mathrm{rel}}$. All values are calculated at ${a_s = 4 \, a_\mathrm{rel}}$ for lattice depths of ${V_1 = V_2 = 10 \, E_\mathrm{r,rel}}$ and a wavelength of ${\lambda = 1 \, \mu}$m. ${E_0 := E_\mathrm{c.m.} + E_\mathrm{rel}}$ is the energy of the uncoupled Hamiltonian. Including $H_{\Delta \omega}$ into the Hamiltonian reduces the energy by ${\Delta E_{\Delta \omega}}$ and including ${H_{\Delta \omega} + V_\mathrm{corr}}$ reduces the energy further by ${\Delta E_\mathrm{corr}}$. The value in brackets is the percentage contribution of the individual coupling terms to the total change of the energy $\Delta E$.}
  \begin{ruledtabular}
    \begin{tabular}{lccccc}
      atom pair & $E_0$ & $\Delta E_{\Delta \omega}$ & $\Delta E_\mathrm{corr}$ & $\Delta E$ \\
      \hline
      $^{40}$K and $^{87}$Rb  & 3.74 & -0.12 (29\%)     & -0.27 (71\%) & -0.39 \\
      $^{6}$Li and $^{133}$Cs & 2.88 & -0.35 (62\%)     & -0.22 (38\%) & -0.57 \\
      $^{6}$Li and $^{87}$Rb  & 2.99 & -0.36 (61\%)     & -0.22 (39\%) & -0.58 \\
      $^{6}$Li and $^{40}$K   & 3.24 & -0.31 (58\%)     & -0.24 (42\%) & -0.55 \\
      $^{6}$Li and $^{7}$Li   & 3.92 & -0.01 $\,$ (2\%) & -0.29 (98\%) & -0.30
    \end{tabular}
  \end{ruledtabular}
\end{table}

Table \ref{tab:table1} shows the effect of the individual coupling terms, ${H_{\Delta \omega} := \mu \Delta\omega^2 \vec r \cdot \vec R}$ and $V_\mathrm{corr}$, on to the energy of several atom pairs. The energies have been calculated for repulsively interacting atoms at ${a_s = 4 \, a_\mathrm{rel}}$ which is the largest scattering length shown in Figs.~\ref{fig:eofa1} and~\ref{fig:eofa2}. All energies of Table \ref{tab:table1} are given in units of the level spacing of the relative motion ${\hbar \omega_\mathrm{rel}}$. Adding the coupling term $H_{\Delta \omega}$ contributes up to 62\% to the total change $\Delta E$ for $^6$Li and $^{133}$Cs. The strong influence of $H_{\Delta \omega}$ stems from the large mass ratio which results in extremely different trap frequencies $\omega_\mathrm{Li}$ and $\omega_\mathrm{Cs}$. By contrast, the energy of $^6$Li and $^7$Li atoms is nearly not affected by $H_{\Delta \omega}$ since the trap frequencies are almost equal.

\section{Experimental procedure}

In the experiment, we have tested our theoretical approach by studying the energy spectrum of $^{40}$K and $^{87}$Rb atom pairs at a single lattice site of a 3D optical lattice in the vicinity of a heteronuclear Feshbach resonance, allowing $a_s$ to be tuned from strong attractive to repulsive interactions. Our experimental procedure for obtaining Fermi-Bose mixtures~\cite{IntDrivDyn,jmo} in optical lattices has been discussed previously~\cite{ImpIndLoc,HeteronuclearMolecules}. In our experiment, we obtain a mixture of $^{40}$K atoms in the ${\left| F=9/2, m_F=9/2 \right>}$ state and $^{87}$Rb in the $\left|F=2,m_F=2\right>$ state by rf-induced sympathetic cooling in a magnetic trap. The mixture is transferred into a crossed optical dipole trap with final trap frequencies for $^{87}$Rb of $2\pi 50$\,Hz. In the optical dipole trap, $^{87}$Rb atoms are transferred from $\left|2,2\right>$ to $\left|1,1\right>$ by a microwave sweep at 20\,G and any remaining atoms in the upper hyperfine $\left|F=2,m_F\right>$ states are removed by a resonant light pulse. Next, we transfer $^{40}$K into the $\left|9/2,-7/2\right>$ state by performing an rf sweep at the same magnetic field with almost 100\% efficiency. With the mixture in the $^{87}$Rb$\left|1,1\right>\otimes^{40}$K$\left|9/2,-7/2\right>$ state, we ramp up the magnetic field to final field values at the Feshbach resonance occurring around 547\,G ~\cite{JILAFesh,LENSFesh,TuningInteractions}. Note that the state which we prepare is not Feshbach-resonant at the magnetic field values which we study, and that a final transfer of $^{40}$K into the $\left|9/2,-9/2\right>$ state is necessary to access the resonantly interacting state. This is precisely the transition which we use to measure the energy spectrum as outlined further below.

We ramp up a 3D optical lattice at a wavelength of 1030\,nm, where the trapping potential for both species is related according to ${V_\mathrm{K}=0.86 \, V_\mathrm{Rb}}$. Due to the different masses of the two species, the trapping frequencies are ${\omega_\mathrm{K}=\sqrt{87/40\cdot0.86} \, \omega_\mathrm{Rb} \simeq 1.4 \, \omega_\mathrm{Rb}}$ in the harmonic approximation. The optical lattice light is derived from a frequency stabilized 20W Yb:YAG disc laser with a 50\,ms linewidth of 20\,kHz. The lattice is formed by three retroreflected laser beams with orthogonal polarizations and a minimum detuning of 10\,MHz between individual beams. In order to get a maximum of lattice sites occupied by one boson and one fermion, the best trade-off has been to limit the particle number at this stage to a few ten thousand.

\begin{figure}[t]
  \centering
  \includegraphics[width=.84\columnwidth]{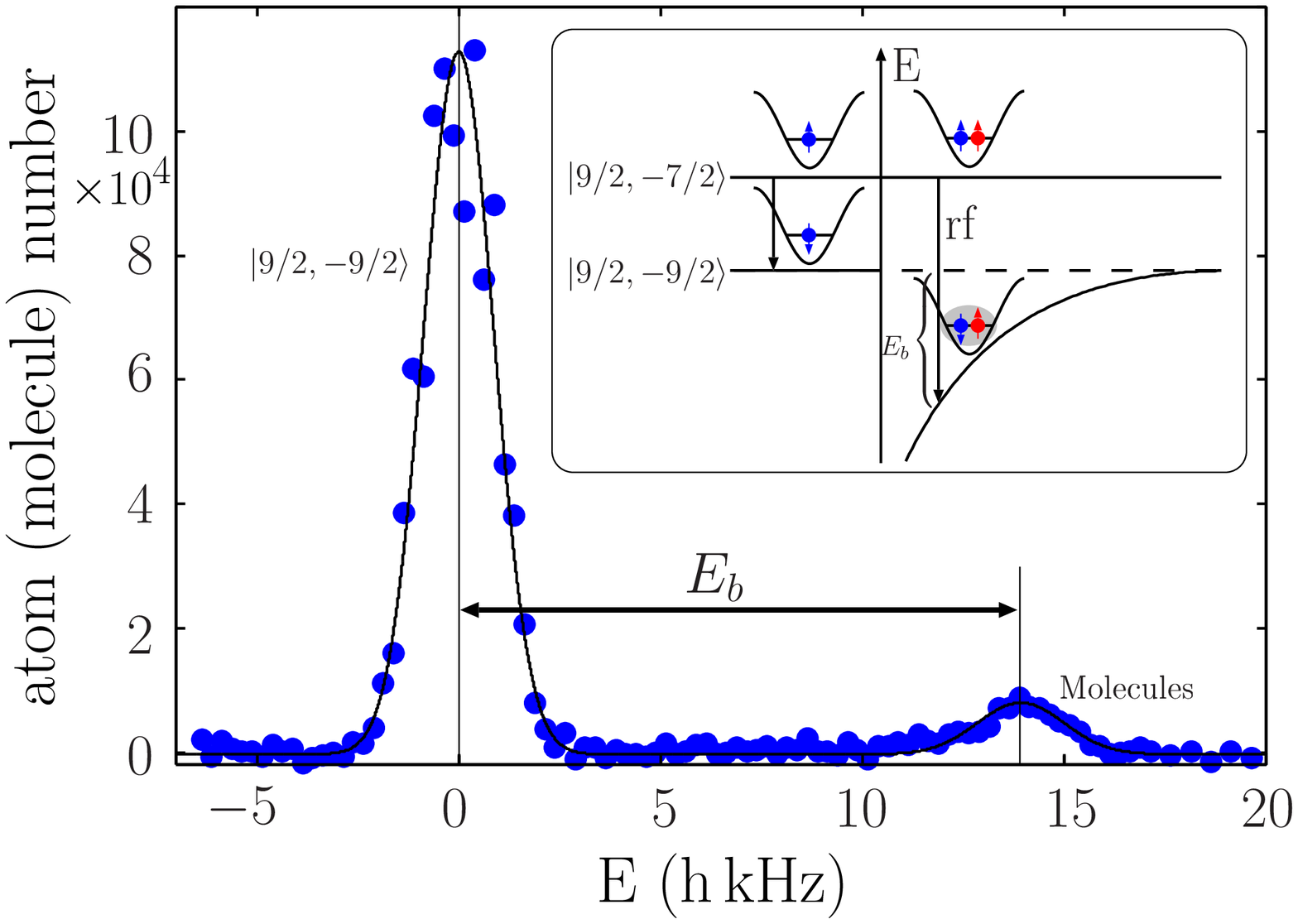}
  \caption{(Color online) rf spectroscopy of $^{40}$K - $^{87}$Rb in a 3D optical lattice on the $^{40}$K${\left| 9/2, -7/2 \right> \rightarrow \left| 9/2, -9/2 \right>}$ transition (see inset) at a lattice depth of ${V_{\mathrm{Rb}} = 27.5 \, E_{r,\mathrm{Rb}}}$ and a magnetic field of 547.13\,G, where the interaction is attractive. The spectrum is plotted as a function of detuning from the undisturbed atomic resonance frequency and clearly shows the large atomic peak at zero detuning. The peak at 13.9\,kHz is due to association of $\left|1,1\right>\otimes\left|9/2,-7/2\right>$ atom pairs into a bound state (figure from Ref.~\cite{HeteronuclearMolecules}).}
  \label{fig:rfspect}
\end{figure}

In the optical lattice, we study the energy spectrum (binding energy) of two particles at a single lattice site by rf spectroscopy (association) of pairs of one $^{87}$Rb and one $^{40}$K atom (see Fig.~\ref{fig:rfspect}). The idea for the measurement is to drive an rf transition between two atomic sublevels one of which is characterized by the presence of the Feshbach resonance and exhibits a large variation of the scattering length as a function of magnetic field according to~\cite{Koehler_review}
\begin{equation} \label{FeshbachFormula}
  a_s(B)=a_\mathrm{bg} \left( 1-\frac{\Delta B}{B-B_0} \right) ,
\end{equation}
where $a_\mathrm{bg}$ is the non-resonant background scattering length, ${\Delta B}$ the magnetic field width of the resonance, and $B_0$ the resonance center position. The other level involved in the rf transition is characterized by a non-resonant scattering length independent of magnetic field over the experimentally studied field range. We use the $^{40}$K $\left|9/2,-7/2\right>\rightarrow\left|9/2,-9/2\right>$ transition where the Feshbach-resonant state is the final $\left|1,1\right>\otimes\left|9/2,-9/2\right>$ state.

A sample spectrum of this transition for  the mixture in the optical lattice is shown in Fig.~\ref{fig:rfspect}. The figure shows two peaks; one of them occurs at the frequency corresponding to the undisturbed $^{40}$K $\left|9/2,-7/2\right>\rightarrow\left|9/2,-9/2\right>$ Zeeman transition frequency at lattice sites occupied by a single $^{40}$K fermion. This peak is used for the calibration of the magnetic field across the Feshbach resonance using the Breit-Rabi formula for $^{40}$K and the hyperfine parameters from Ref.~\cite{HyperfineAlkalis}. For 57 measurements on 11 consecutive days, we find a mean deviation from the magnetic field calibration of 2.7\,mG at magnetic fields around 547\,G, corresponding to a field reproducibility of $5 \times 10^{-6}$. There is an additional uncertainty on the absolute value of the magnetic field due to the specified reference frequency source accuracy for the rf synthesizer of $1 \times 10^{-5}$, resulting in an uncertainty of the measured magnetic fields of 12\,mG.

The second peak at a positive detuning of 13.9\,kHz is the result of interactions between $^{40}$K and $^{87}$Rb at a lattice site where one heteronuclear atom pair is present. There are two different energy shifts causing the observed separation of the peaks: One is the constant, small energy shift of the initial $\left|1,1\right>\otimes\left|9/2,-7/2\right>$ state which is independent of $B$, and the important, magnetic field sensitive collisional shift which stems from the strong Feshbach-resonant interactions in the $\left|1,1\right>\otimes\left|9/2,-9/2\right>$ final state. In the specific example, the negative energy shift (binding energy) of the final state increases the transition frequency as seen in Fig.~\ref{fig:rfspect}. In order to perform spectroscopy on the aforementioned transition, we use pulses with a gaussian amplitude envelope ($1/e^2$ full width of 400\,$\mu$s and total pulse length of 800\,$\mu$s), resulting in an rf $1/e^2$ half linewidth of 1.7\,kHz. We choose the pulse power such as to achieve full transfer on the single atom transition, i.\ e.\ rf pulse parameters including power are identical for all magnetic fields. Not only for rf spectroscopy pulse generation, but also for evaporation and state transfer, we have used an advanced rf synthesizer\footnote{VFG-150, a FPGA-driven fast Versatile-Frequency-Generator, controlled via an USB2.0 interface and developed by Th. Hannemann in the group of C. Wunderlich, University of Siegen, Germany.} allowing precise control of frequency, amplitude and phase down to the 5\,ns level and therefore the synthesis of in principle arbitrary pulse shapes.

\section{Experimental \lowercase{vs} theoretical spectrum. Resonance position}

From rf spectra as in Fig.~\ref{fig:rfspect}, we can determine the separation between the single atom and the two-particle (``molecular'') peak with high precision (typical uncertainty of 250\,Hz) and thus extract the binding energy up to a constant offset due to non-zero background scattering lengths. At the same time, the atomic peak provides us with a precise magnetic field calibration as described above. Spectra as in Fig.~\ref{fig:rfspect} have been recorded for magnetic field values across the whole resonance and yield the energy spectrum as a function of magnetic field.

\begin{figure}[t]
  \centering
  \includegraphics[width=.86\columnwidth]{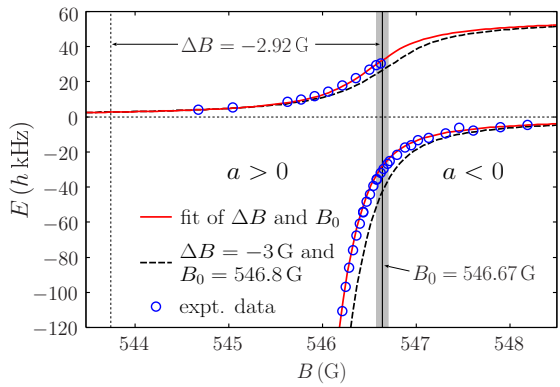}
  \caption{(Color online) Experimentally observed energy spectrum together with theory without free parameters (black dashed line) and a least squares fit for the resonance parameters $B_0$ and ${\Delta B}$ (red solid line). Experimental data from Ref.~\cite{HeteronuclearMolecules}.}
\label{fig:eofb}
\end{figure}

Fig.~\ref{fig:eofb} shows the measured energy shift across the resonance at a lattice depth of 40.5$E_{r,\mathrm{Rb}}$ as a function of magnetic field. The energy shift is obtained from Fig.~\ref{fig:eofa1} by subtracting the energy of the initial $^{87}$Rb$\left|1,1\right>\otimes^{40}$K$\left|9/2,-7/2\right>$ state: ${E_s = E - E(a_{-7/2} = -175 a_0)}$~\cite{a_initial}. In addition, Figs.~\ref{fig:eofb} and~\ref{fig:eofa1} are connected through Eq.~(\ref{FeshbachFormula}). One branch of the spectrum is characterized by the presence of a ``positive'' binding energy, the repulsively interacting pair branch. In Fig.~\ref{fig:eofa1}, we have seen that this branch continuously transforms into attractively interacting atoms as a function of $a_s$. As a function of magnetic field, however, and as a result of Eq.~(\ref{FeshbachFormula}), we observe this transition as a jump from the left-hand side of Fig.~\ref{fig:eofb}, where the interaction is weak and repulsive, to the right-hand side of Fig.~\ref{fig:eofb}, where the interaction is weak and attractive. Here, we find attractively interacting atoms which decay into free atom pairs if the external confinement of the optical lattice is removed.

Whereas in Fig.~\ref{fig:eofa1}, the attractively interacting atoms branch and the molecule branch are only asymptotically equal in the limit $\left|a_s\right|\rightarrow\infty$, the singularity on resonance in Eq.~(\ref{FeshbachFormula}) transforms this into a continuous crossover across the center of the resonance position as a function of magnetic field and as seen in Fig.~\ref{fig:eofb}. These molecules are stable even in the absence of the optical lattice potential.

In order to compare the numerically calculated energy spectrum ${E(a_s)}$ (blue solid line of Fig.~\ref{fig:eofa1}) to the experimental data ${E(B)}$ of Fig.~\ref{fig:eofb}, we transform the scattering length $a_s$ into the magnetic field strength $B$ via Eq.~(\ref{FeshbachFormula}). By using parameters from the literature: ${a_\mathrm{bg} = -185 a_0}$, ${\Delta B = -3}$\,G~\cite{ControlInteractions} and ${B_0 = 546.8 }$\,G~\cite{TuningInteractions}, we obtain the black dashed curve in Fig.~\ref{fig:eofb}. As can be seen, the difference between the theoretical prediction and the experimental data can be overcome by a shift of the black dashed curve along the magnetic field axis. We attribute this shift to an insufficient knowledge of the resonance center position $B_0$.

We therefore fit our theoretical calculations to the experimental data in order to improve the estimate for the resonance center position $B_0$. As independent fit parameters we choose $B_0$ and ${\Delta B}$, while $a_\mathrm{bg}$ is fixed. The latter parameter cannot be determined independently from the measurements due to its strong correlations with ${\Delta B}$. This is due to the fact that in the vicinity of the resonance center position $B_0$ the first term of Eq.~(\ref{FeshbachFormula}) is negligible so that only the product ${a_\mathrm{bg} \Delta B}$ can be determined precisely from the fit. We therefore set ${a_\mathrm{bg} = -185 a_0}$~\cite{ControlInteractions} and use ${\Delta B}$ and $B_0$ as free fit parameters, with the caveat that only the value obtained for $B_0$ is to be considered precise. In Fig.~\ref{fig:eofb}, the result of the least squares fit is displayed as a red solid line. Note that the reliability of the fitting procedure sensitively depends on an accurate calculation of the energy spectrum ${E(a_s)}$ which includes an exact treatment of the anharmonicity and the different trap frequencies of the two atoms.

The least squares fit results in the following values of the resonance parameters ${\Delta B = -2.92}$\,G and ${B_0 = 546.669}$\,G. The fit results in an uncertainty of 2\,mG on $B_0$. The value of $B_0$ sensitively depends on the scattering length of the initial state $a_{-7/2}$. Assuming an uncertainty of $a_{-7/2}$ of 10\% results in an uncertainty on $B_0$ of 20\,mG. Another possible source of systematic uncertainties may be the lattice depth calibration. The lattice depth has been calibrated by parametric excitation from the first to the third band of the lattice and is estimated to have an uncertainty of 5\%. Repeating the fit procedure with $\pm$5\% variations on the lattice depth calibration, we obtain a corresponding systematic uncertainty on $B_0$ of 7\,mG. A third source of systematic uncertainties finally stems from the finite basis and an imprecise approximation of the lattice site potential. Here, we included corrections up to eighth order and generated basis states of the lowest eight energy levels of the uncoupled Hamiltonian. This improved the value of $B_0$ by 2\,mG compared to a calculation with up to sixth order corrections and basis states of lowest seven energy levels. Adding the systematic uncertainty of the magnetic field calibration of 12\,mG (see above), we finally obtain
\begin{equation} \label{B0_ergebnis}
B_0=546.669(24)_{\mathrm{syst}}(2)_{\mathrm{stat}}\,\mathrm{G}
\end{equation}
under the assumption that the pseudopotential treatment is valid in the present experimental situation~\cite{validity}.

\vspace{-3ex}

\section{Efficiency of \lowercase{rf} association}

\vspace{-1.5ex}

In a next step, we have analyzed the transfer efficiency of the rf association. The rf association process can be described theoretically by a Rabi model: The spin of the $^{40}$K atoms is flipped from ${| 9/2, -7/2 \rangle}$ to ${| 9/2, -9/2 \rangle}$ by applying a radio frequency. The atoms are initially in state ${|1\rangle := (\Phi_i,0)}$ and afterwards in state ${|2\rangle := (0,\Phi_f)}$. In the rotating frame and by integrating out the spatial degrees of freedom we obtain the Hamiltonian
\begin{equation} \label{Hrf}
H_\mathrm{rf} = \frac{\hbar}{2} \begin{pmatrix} -\Delta \omega & \langle \Phi_i | \Phi_f \rangle \omega_1(t) \\ \langle \Phi_i | \Phi_f \rangle \omega_1(t) & +\Delta \omega \end{pmatrix} \; ,
\end{equation}
where ${\Delta \omega := \omega - \omega_0 - \omega_b}$ is the detuning, $\omega$ is the radio frequency, ${\omega_0 \propto B_0}$ is proportional to the applied magnetic field, $\omega_b$ is proportional to the binding energy ${E_b = \hbar \omega_b}$, ${\omega_1(t)}$ is proportional to the amplitude of the oscillating magnetic field ${B_1(t)}$, and ${\langle \Phi_i | \Phi_f \rangle}$ is the overlap integral between the initial and final motional wave functions. As can be seen, the off-diagonal elements of Hamiltonian (\ref{Hrf}) are not only proportional to the rf amplitude ${\omega_1(t)}$, but also to the overlap integral ${\langle \Phi_i | \Phi_f \rangle}$. Therefore, the transfer probability between states ${|1\rangle}$ and ${|2\rangle}$ corresponds to Rabi flopping with a Rabi frequency reduced by the overlap integral of $\Phi_i$ and $\Phi_f$ compared to the pure atomic transition. Exactly on the molecular resonance, we have ${\omega = \omega_0 + \omega_b}$ (${\rightarrow \Delta \omega = 0}$). The on-resonant result for the theoretical transfer probability (efficiency) is thus given by
\begin{equation} \label{teff}
P_{1 \rightarrow 2} = \sin^2\left[ \frac{1}{2} \langle \Phi_i | \Phi_f \rangle \int_0^t \omega_1(t') dt' \right]
\end{equation}
which is unity for a transfer between atomic states, where ${\Phi_i = \Phi_f}$, when setting the area under the $\omega_1(t)$ curve to ${\int_0^t \omega_1(t') dt' = \pi}$. For transfer into the molecular state, the theoretical probability decreases as a function of the wave function overlap integral since the molecular final orbital wave function becomes more and more dissimilar from the initial two-body atomic wave function.

In the experiment, the molecules were associated using rf pulses designed to induce a $\pi$ pulse for the non-interacting atoms: $\int_0^t\omega_1(t')\,dt'=\pi$. This $\pi$ pulse has been kept fixed over the entire range of magnetic field values investigated. The experimental association efficiency is determined from the height of the molecular peak (see Fig. 3) as a function of magnetic field for constant pulse parameters and $\omega=\omega_0+\omega_b$ ($\rightarrow\Delta \omega=0$) as in the theory above.

\begin{figure}[t]
  \centering
  \includegraphics[width=.85\columnwidth]{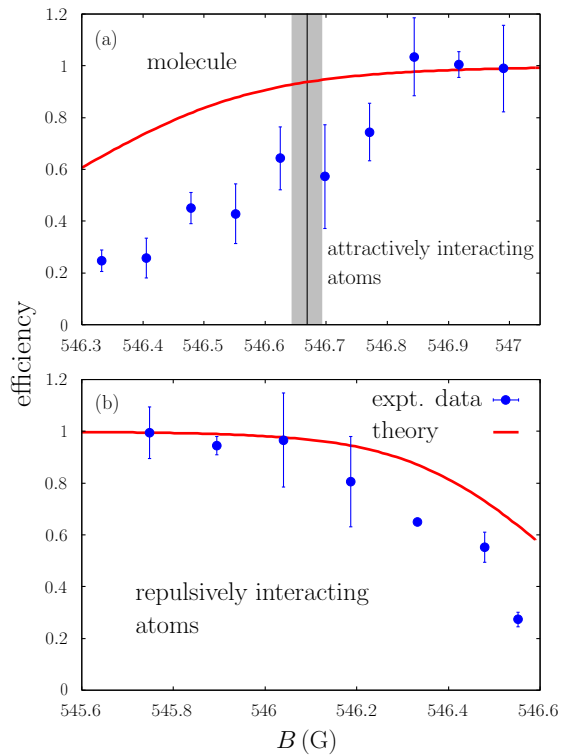}
  \caption{(Color online) Transfer efficiency of rf association as observed in the experiment and estimated from a Rabi model, both for (a) attractively interacting atoms and molecules, and for (b) repulsively interacting pairs. The experimental data contain a global factor which has been chosen such that the value of the outermost right (a) [left (b)] data point is one (see text). Expt. data of (a) from Ref.~\cite{HeteronuclearMolecules}.}
  \label{fig:teff}
\end{figure}

Figs.~\ref{fig:teff}(a) and \ref{fig:teff}(b) show a comparison between the conversion efficiency as extracted from the experimental data and the theoretical estimate from equation (\ref{teff}). Theory and experiment show the general trend of dropping association efficiency with increasing binding energy when the initial and final wave functions become more and more dissimilar. In this context, we define the experimental conversion efficiency as the ratio of the number of molecules created and the initial lattice sites which are occupied by exactly one K and one Rb atom. Note that only on these lattice sites molecules can be created. For the comparison of experimental and theoretical transfer efficiency, the experimental data have been scaled by a global factor to reproduce a conversion efficiency of 1 far off the Feshbach resonance where initial and final two-body wave function are equal. This is necessary, because the initial lattice sites occupied by one K and one Rb atom have not been determined experimentally.

While the experiments presented here were performed at constant rf pulse parameters, it should be possible from the above arguments to increase either pulse power or duration or both  of the rf pulse to account for the reduced wave function overlap and thereby always obtain an efficiency of 1. In particular, it should be possible to drive Rabi oscillations between atoms and molecules in a very similar way as recently demonstrated~\cite{mpq_oscillations}. The comparison also indicates that in the case of association efficiency a full quantitative agreement might require some more sophisticated treatment of the association process. This is in contrast to the analysis of binding energies and lifetimes (see below), where the good quantitiative agreement shows that here the $\delta$ interaction approximation captures the essential physics. Testing the Rabi oscillation hypothesis for molecules with rf might provide further insight.

\vspace{-3ex}

\section{Lifetime}

\vspace{-1.5ex}

Molecule formation at atomic Feshbach resonances results in dimers which are very weakly bound and may exhibit strong inelastic collisional losses. Experiments with molecules created from bosonic atoms showed very small lifetimes. As a result, these molecules can be brought into the quantum degenerate regime~\cite{QuantDegMol}, but thermal equilibrium is generally difficult to achieve for molecules created from bosonic atoms because of the short lifetime.

In experiments with molecule creation from two-component Fermi gases~\cite{MoleculeCreationFermi}, inelastic molecule-molecule and molecule-atom collisions are suppressed by the Pauli exclusion principle~\cite{PetrovDimers,PetrovDimers_2}, resulting in remarkably long lifetimes between approximately 100~ms and even seconds, allowing Bose-Einstein condensation of Feshbach molecules~\cite{MoleculeCondensation} and the observation of BCS-BEC crossover physics~\cite{bcsbec1,bcsbec2,bcsbec3,bcsbec4,bcsbec5}. The lifetime limitation for molecules created from bosonic atoms has been overcome by creating molecules in 3D optical lattices, where molecules are created at a single lattice site and isolated from inelastic collisions with residual atoms or other molecules~\cite{LattMolBose}.

For molecules created from Fermi-Bose mixtures, the situation is a little bit more complicated. As far as collisions between molecules are concerned, the fermionic character of the molecule should become more evident the deeper the molecule is bound, thus resulting in suppression of collisions~\cite{MITFesh}.

As far as collisions with residual atoms are concerned, we expect inelastic collisions with fermionic atoms in the same spin state as the fermionic component of the molecule, i.\ e.\ in the $\left|9/2,-9/2\right>$ state, to be suppressed due to the Pauli exclusion principle close to the resonance, when the ``atomic'' character of the molecule's constituents is still significant~\cite{PetrovDimers,PetrovDimers_2}. For collisions with bosonic atoms and fermionic atoms in a different internal state, we do not expect any Pauli-blocking enhanced lifetime, since the residual atom can in principle come arbitrarily near to the molecule's constituents.

In our situation, where the molecules are created through rf association, residual fermionic atoms remain in a different spin state, either in ${\left|9/2,-7/2\right>}$ or ${\left|9/2,-5/2\right>}$ (for the latter case, and for a description of the experimental procedure, see Ref.~\cite{HeteronuclearMolecules}). These residual fermionic atoms as well as the remaining bosonic atoms may therefore limit the stability of the molecular sample.

\begin{figure}[t]
  \centering
  \includegraphics[width=\columnwidth]{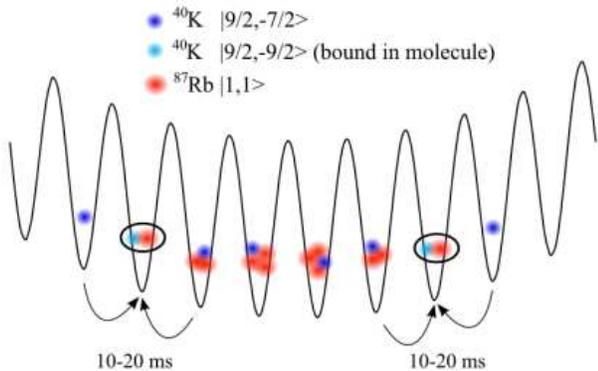}
  \caption{(Color online) Sketch of expected lattice occupation. The arrows illustrate tunneling of remaining fermionic $^{40}$K atoms to the ``molecular'' shell where they can undergo inelastic three-body collisions with a $^{40}$K-$^{87}$Rb molecule.}
  \label{fig:lattice_occupation}
\end{figure}

\begin{figure}[t]
  \centering
  \includegraphics[width=.9\columnwidth]{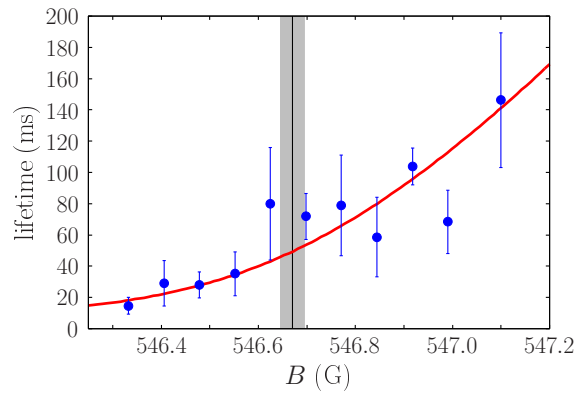}
  \caption{(Color online) Lifetime of heteronuclear $^{40}$K-$^{87}$Rb molecules in the optical lattice. The Lifetime is limited due to residual atoms which can tunnel to lattice sites with molecules and provoke inelastic three-body loss. The theoretical prediction uses the pseudopotential wave function and contains a global factor which was adjusted to the experimental data of Ref.~\cite{HeteronuclearMolecules}.}
  \label{fig:lifetime}
\end{figure}

Molecule creation in the optical lattice introduces a second aspect concerning the lifetime: lattice occupation and tunneling probabilities. In Fig.~\ref{fig:lattice_occupation}, we have sketched the expected occupation in the optical lattice. Prior to molecule creation, we expect slightly less than unity filling for the fermionic component. As far as the bosons are concerned, we expect a central occupation number between 3 and 5, surrounded by shells of decreasing occupation number. In the rf association process, molecules are only created in the shell where we have one fermion and one boson per lattice site. In the outermost region of the lattice, we have lattice sites with only one fermion which are responsible for the ``atomic'' peak in the rf spectroscopy signal. After the rf association process in the ``molecular'' shell, bosons from neighboring sites as well as fermions remaining in a different spin state can tunnel to the ``molecular'' shell and provoke inelastic three-body loss. In our experimental situation, this is more probable for the remaining fermionic atoms, since they are lighter and have a smaller tunneling time (10 to 20 ms for the lattice depths discussed here). For the highest binding energies observed in the experiment, we find a limiting lifetime of 10 to 20\,ms as seen in Fig.~\ref{fig:lifetime}, which is consistent with the assumption that in this case, three-body loss is highly probable once tunneling of a distinguishable residual fermion has occured. Still, for the more weakly bound molecules and in particular for attractively interacting atoms, we observe high lifetimes of 120\,ms, raising the question of the magnetic field dependence of the lifetime.

We can understand this magnetic field dependence using the pseudopotential model by introducing a product wave function for the combined wave function of the resonantly interacting atom pair and a residual fermionic atom after tunneling to a molecular site. We write this three-body wave function as
\begin{equation} \label{PPsol}
  \Phi(\vec r, \vec R, \vec r_3) := \Phi_\mathrm{mol}(\vec r, \vec R) \Phi_3(\vec r_3)
\end{equation}
where $\Phi_\mathrm{mol}$ is the result of the pseudopotential calculation for the molecule and $\Phi_3$ is the ground-state wave function of the residual atom at the same lattice site. Note that this treatment assumes weak interactions between the residual atom and the molecule (the interaction between the residual atom and the molecule's constituents is on the order of the background scattering length). From solution (\ref{PPsol}) of the pseudopotential model, the dependence of the loss rate on the scattering length can be obtained close to the resonance~\cite{PetrovDimers}: the loss rate $\Gamma$ is proportional to the probability $\mathcal{P}$ of finding the three atoms within a small sphere of radius $\sigma$, where they can undergo three-body recombination. This probability is expected to become larger for more deeply bound molecules, since two of the three atoms are already at a close distance. Up to a global factor, $\mathcal{P}$ is independent of the value chosen for $\sigma$, provided ${\sigma \ll a_\mathrm{rel}}$, and also $\sigma\ll a_s$ in the molecule regime. More quantitatively, we calculate this probability according to
\begin{equation} \label{probability}
\mathcal{P}=\int_{|\vec r|<\sigma \atop |\vec r_3 - \vec R| < \sigma} d\vec r\,d\vec R\,d\vec r_3 \bigl|\Phi(\vec r, \vec R, \vec r_3)\bigr|^2 .
\end{equation}
The magnetic field dependence of the loss rate is thus given through $\Gamma\propto\mathcal{P}$, and the lifetime is proportional to $1/\Gamma$. By using the wave functions (\ref{phi_busch}) and the relation ${|A|^2\propto a_s^2 \, d E_\mathrm{rel} / d a_s}$~\cite{Busch98}, we obtain
\begin{equation} \label{loss_rate}
\mathcal{P} = C \, \frac{a_s}{\psi(-\frac{E_\mathrm{rel}}{2\hbar \omega_\mathrm{rel}} + \frac{3}{4}) - \psi(-\frac{E_\mathrm{rel}}{2\hbar \omega_\mathrm{rel}} + \frac{1}{4})} \, ,
\end{equation}
where $C$ is independent of $a_s$, and ${\psi(x)\equiv \Gamma'(x)/\Gamma(x)}$ is the digamma function. This result agrees with a numerical integration of Eq.~(\ref{probability}) using the eigenfunctions of the complete Hamiltonian (\ref{H}).

The lifetime obtained from the calculation is shown in Fig.~\ref{fig:lifetime} as a red solid line, scaled by a global factor to allow comparison to the experiment. As can be seen, the theoretical prediction explains the magnetic field dependence of the lifetime rather well. From an experimental point of view, we can therefore expect that removal of the remaining atoms using a resonant light pulse will significantly increase the lifetime of the molecules in the optical lattice.

\vspace{-2ex}

\section*{\uppercase{Conclusions}}

\vspace{-2ex}

To summarize, we have developed a pseudopotential approach to the scattering of unequally trapped atoms at a single site of an optical lattice including terms which couple center-of-mass and relative motion. We have compared the energy spectrum to experimental results for $^{40}$K and $^{87}$Rb atoms interacting at a heteronuclear Feshbach resonance in a 3D optical lattice. Within the pseudopotential model we have precisely determined the center position of the Feshbach resonance based on this comparison. The pseudopotential approach also allows us to understand the efficiency of rf association used to experimentally determine the energy spectrum, as well as the dependence of the molecular lifetime on magnetic field. The model developed in this paper enables a broad understanding of heteronuclear atom pairs in an optical lattice. We are aware of possible limitations of the pseudopotential model. It might be an interesting option to extend the method described here to energy dependent pseudopotentials~\cite{Greene,Bolda,Petrov_bosons,Idziaszek,Naidon,Pricoupenko} or multi-channel models~\cite{Bolda,Tiesinga,Dickerscheid}. Finally, we note that the present rf association technique could be used to study the three-body problem at a triply occupied lattice site~\cite{Stoll_Koehler,Felix}. An advantage of this method with respect to the adiabatic magnetic field sweep proposed in~\cite{Stoll_Koehler} is that it is less sensitive to three-body losses, which are particularly important for Efimovian states~\cite{Felix}.

\vspace{-4ex}

\section*{\uppercase{acknowledgments}}

\vspace{-2ex}

The authors would like to thank K.~Rz{\c a}{\.z}ewski for his contributions in the initial discussion of possible extensions to Ref.~\cite{Busch98}. We acknowledge discussions with Y.~Castin, P.~Julienne, Th.~K\"ohler, P.~Naidon, D.~Petrov, and L.~Pricoupenko on the pseudopotential models and on loss rates. We would like to thank A.~Simoni for providing us with the closed channel magnetic moment at the Feshbach resonance and the scattering length in the initial state of the rf association. We acknowledge financial support by the Deutsche Forschungsgemeinschaft (SPP 1116).


\begin{thebibliography}{22}
\bibitem{validity}
We expect the pseudopotential model to be fairly accurate for the experimental parameters of~\cite{HeteronuclearMolecules}. Indeed, even for large scattering lengths, this model is expected to become exact in the zero-range limit~\cite{Castin}, that is when $1 / k_\mathrm{typ} \gg \max(\beta_6,|r_e^0|)$. Here, $k_\mathrm{typ}$ is the typical wavenumber of the relative motion of the two atoms, ${\beta_6 = (2\mu C_6/\hbar^2)^{1/4}}$  is the van der Waals length scale, and $r_e^0 := -\hbar^2/(\mu a_\mathrm{bg} \Delta B \partial E_\mathrm{res} / \partial B)$, ${\partial E_\mathrm{res}/ \partial B}$ being the magnetic moment of the closed channel with respect to the two-atom open channel. For K-Rb in their ground state, ${\beta_6 = 7.6 \, \text{nm}}$~\cite{C6}. Using ${\partial E_\mathrm{res}/ \partial B = k_B 144 \, \mu \text{K/G}}$~\cite{SimoniPrivComm}, we get ${r_e^0 = -4.6 \, \text{nm}}$. It remains to estimate $k_\mathrm{typ}$. In the molecule regime, we have ${1/k_{\rm typ}\sim a_s > 47 \, \text{nm}}$. In the other regimes (attractively and repulsively interacting atoms), we have ${1/k_\mathrm{typ}\sim a_{\rm rel}}$. In the harmonic approximation for the experimental lattice depth, ${a_\mathrm{rel} = 103 \, \text{nm}}$. Thus, the above inequality is fairly well satisfied.
\bibitem{Basis}
We note that we neglect many low-energy basis states ${|N,L,0,0,0,0\rangle}$ with ${2N+L > \Pi_\mathrm{max}}$ when the scattering length is small and positive ${a_s \gtrapprox 0}$. These molecule states with highly excited center-of-mass energy have nevertheless low total energy since the energy of the relative motion is large and negative. We have included the states ${|N,L,0,0,0,0\rangle}$ with ${2N+L \leq \Pi_\mathrm{max}}$ to allow for the flattening of the center-of-mass wave function due to the anharmonicity of the lattice site potential. But we have neglected the states with ${2N+L > \Pi_\mathrm{max}}$ since they are only weakly coupled to the states ${|0,0,0,0,0,0\rangle}$ (molecule) and ${|0,0,0,n = 1,0,0\rangle}$ (repulsively interacting atoms of the decoupled problem). These matrix elements become even zero if ${2N+L > p}$ (${p = 1, ..., 8}$) since in this case the matrix elements ${\langle 0,0,0,0,0,0| X^p |N,L,0,0,0,0\rangle}$ and ${\langle 0,0,0,1,0,0| X^p |N,L,0,0,0,0\rangle}$ are exactly zero.
\end{thebibliography}
\end{document}